\documentclass[preprint]{aastex}

\begin{document}

\title {The Eclipsing $\delta$ Scuti Star EPIC 245932119 }
\author{Jae Woo Lee$^{1,2}$, Kyeongsoo Hong$^{1,3}$, and Martti H. Kristiansen$^{4,5}$  }
\affil{$^1$Korea Astronomy and Space Science Institute, Daejeon 34055, Korea}
\affil{$^2$Astronomy and Space Science Major, Korea University of Science and Technology, Daejeon 34113, Korea}
\affil{$^3$Institute for Astrophysics, Chungbuk National University, Cheongju 28644, Korea}
\affil{$^4$DTU Space, National Space Institute, Technical University of Denmark, Elektrovej 327, DK-2800 Lyngby, Denmark}
\affil{$^5$Brorfelde Observatory, Observator Gyldenkernes Vej 7, DK-4340 T\o{}ll\o{}se, Denmark}
\email{jwlee@kasi.re.kr}

\begin{abstract}
We present the physical properties of EPIC 245932119 ($K_{\rm p}$ = $+$9.82) exhibiting both eclipses and 
pulsations from the {\it K{\rm 2}} photometry. The binary modeling indicates that the eclipsing system is in detached 
or semi-detached configurations with a mass ratio of 0.283 or 0.245, respectively, and 
that its light-curve parameters are almost unaffected by pulsations. Multiple frequency analyses were 
performed for the light residuals in the outside-primary eclipsing phase after subtracting the binarity effects from 
the observed data. We detected 35 frequencies with signal to noise amplitude ratios larger than 4.0 in two regions of 
0.62$-$6.28 day$^{-1}$ and 19.36$-$24.07 day$^{-1}$. Among these, it is possible that some high signals close to 
the Nyquist limit $f_{\rm Ny}$ may be reflections of real pulsation frequencies (2$f_{\rm Ny}-f_i$). All frequencies 
($f_8$, $f_9$, $f_{14}$, $f_{18}$, $f_{24}$, $f_{32}$) in the lower frequency region are orbital harmonics, and 
three high frequencies ($f_{19}$, $f_{20}$, $f_{22}$) appear to be sidelobes split from the main frequency of 
$f_1$ = 22.77503 day$^{-1}$. Most of them are thought to be alias effects caused by the orbital frequency. 
For the 26 other frequencies, the pulsation periods and pulsation constants are in the ranges of 0.041$-$0.052 days 
and 0.013$-$0.016 days, respectively. These values and the position in the Hertzsprung-Russell diagram reveal that 
the primary component is a $\delta$ Sct pulsator. The observational properties of EPIC 245932119 are in good agreement 
with those for eclipsing binaries with $\delta$ Sct-type pulsating components. 
\end{abstract} 

{}

\section{INTRODUCTION}

For eclipsing binaries (EBs), it is possible to measure precisely and directly fundamental stellar parameters, such as mass, 
radius, and luminosity through time-series photometry and spectroscopy, while pulsating stars provide valuable information 
about their interior structure through asteroseismology. Thus, EBs with pulsating components are attractive objects that 
test stellar structure and evolution models from their binarity and pulsation features. Over 90 of them have been 
known to contain $\delta$ Sct-type pulsating components (Kahraman Ali\c cavu\c s et al. 2017; Liakos \& Niarchos 2017). 
The $\delta$ Sct pulsations are low-order pressure ($p$) modes driven by the $\kappa$ mechanism acting in 
the He II ionization region and assist in probing the envelope of a star. The $\delta$ Sct components in binaries have 
pulsation characteristics similar to single $\delta$ Sct stars, but their pulsations may be influenced by the binary effects, 
such as tidal interaction and mass transfer between both components. Recently, Liakos \& Niarchos (2015, 2017) proposed that 
there is a threshold in the binary period of $\sim$ 13 days below which the binarity affects the pulsation properties and 
there exists a linear correlation between orbital and pulsation periods. Kahraman Ali\c cavu\c s et al. (2017) compared 
the physical parameters of pulsating EBs with those of single pulsators and showed that the $\delta$ Sct components in EBs 
pulsate with shorter periods and lower amplitudes. However, most of the eclipsing $\delta$ Sct stars still have not been 
studied in detail because of the lack of observed data (Liakos \& Niarchos 2017; Kahraman Ali\c cavu\c s et al. 2017). 

This paper is the seventh contribution in a series of detailed studies for the pulsating EBs using the precise photometric data 
from the space missions (Lee et al. 2014, 2016a,b, 2017; Lee 2016; Lee \& Park 2018). We choose the {\it K{\rm 2}} target 
EPIC 245932119 (HD 220687, ASAS J232548-1136.6, RAVE J232547.8-113636, TYC 5825-1038-1; $K_{\rm p}$ = $+$9.820; $V\rm_T$ = $+$9.622, 
$(B-V)\rm_T$ = $+$0.201) with the {\it Kepler} spacecraft (Koch et al. 2010). From the All Sky Automated Survey (ASAS)-3 
database, Pigulski \& Michalska (2007) claimed that the stellar system is a pulsating EB with an orbital period of 
1.594251 $\pm$ 0.000003 days and a dominant frequency of 26.16925 $\pm$0.00004 days$^{-1}$, corresponding to a pulsation period 
of about 0.03821 days. Wraight et al. (2011) identified the variability of this star as an EB with a period of 
1.59430 days and a primary eclipse depth of 0.15 mag using the STEREO observations (Kaiser et al. 2008). On the other hand, 
Gray et al. (2017) conducted a survey in the southern hemisphere to increase the number of $\lambda$ Boo stars, which are 
a rare class of Population I metal-weak A-type stars. They reported that the infrared excesses detected in EPIC 245932119 are 
not the $\lambda$ Boo phenomena but arise from the presence of the cooler component. 

Our results from both the binary modeling and the pulsation analysis indicate that EPIC 245932119 is a short-period EB 
with a $\delta$ Sct-type pulsating component. This study follows the following structure. In section 2, we present 
the light-curve synthesis for the observed {\it K{\rm 2}} data with an extensive mass-ratio ($q$) search. Section 3 
describes the multiple frequency analyses for the eclipse-subtracted light residuals, using the iteration method applied 
in a paper of Lee et al. (2017). Finally, the discussion and conclusions for this work are given in section 4.

\section{{\it K{\rm 2}} PHOTOMETRY AND LIGHT-CURVE SYNTHESIS}

EPIC 245932119 was observed in a long cadence mode with an integration time of 29.42 min during Campaign 12 of 
the ${\it Kepler}$ {\it K{\rm 2}} mission (Howell et al. 2014). We used the simple aperture photometry (SAP) data from 
the MAST Archive\footnote{http://archive.stsci.edu/k2/}. The raw data were detrended by fitting the out-of-eclipse part of 
the orbital light curve to a straight line and were converted to a magnitude scale by requiring a {\it Kepler} magnitude 
of +9.82 at maximum light. As shown in Figure 1, the {\it K{\rm 2}} light curve displays an ellipsoidal variation 
outside eclipses, which may be partly caused by tidal distortion. The depth difference between the primary and secondary 
eclipses indicates a large temperature difference between both components. In order to obtain the light-curve parameters 
of the binary star, we analyzed the {\it K{\rm 2}} data using the 2007 version of the Wilson-Devinney synthesis code 
(Wilson \& Devinney 1971, van Hamme \& Wilson 2007; hereafter W-D). 

The light-curve modeling of our program target was done in a manner similar to that for the pulsating EBs KIC 6220497 
(Lee et al. 2016a) and KIC 11401845 (Lee et al. 2017). The surface temperature of the hotter primary component was set 
to be $T_{1}$ = 7,652 $\pm$ 95 K from the RAVE (RAdial Velocity Experiment; Kordopatis et al. 2013) Catalogue. 
The logarithmic bolometric ($X_{1,2}$) and monochromatic ($x_{1,2}$) limb-darkening coefficients were interpolated from 
the tables of van Hamme (1993). The gravity-darkening exponents ($g_{1,2}$) and the bolometric albedos ($A_{1,2}$) were 
held fixed at standard values of $g_1$ = 1.0 and $g_2$ = 0.32, and $A_1$ = 1.0 and $A_2$ = 0.5, as surmised from 
the components' temperatures. In Figure 1, the secondary eclipses are consistent with one half period after 
the primary eclipses, which implies that EPIC 245932119 has negligible eccentricity. As expected for short-period binaries, 
a synchronous rotation for both components ($F_{1,2}$ = 1.0) was adopted, and the detailed reflection treatment was applied. 
It is known that the $F$ values cause little impact on the light-curve parameters (e.g., Lee et al. 2018). In this paper, 
the adjustable parameters were the orbital ephemeris ($T_0$ and $P$), the inclination angle ($i$), the temperature ($T_2$) 
of the secondary star, the dimensionless surface potential ($\Omega_{1,2}$), and the monochromatic luminosity ($L_{1}$).  

The mass ratio ($q$) is one of the most important parameters in studying the Roche-geometry configuration and physical 
properties of EBs. Nonetheless, neither light-curve solution nor spectroscopic $q$ have been made for EPIC 245932119. 
Thus, we conducted the so-called $q$-search procedure that computes a series of models with varying $q$ for 
various modes of the W-D code. The behavior of the weighted sum of the squared residuals ($\sum{W(O-C)^2}$; hereafter $\sum$) 
was used to estimate the potential reality of each model. The results are displayed in Figure 2, and we found two possible 
Roche geometries: a detached configuration with $q=0.28$ and a semi-detached configuration (the secondary component filling 
its limiting lobe) with $q=0.24$. In the subsequent calculations, the $q$ values were used as an additional free parameter 
in each configuration to derive the binary parameters listed in Table 1. The photometric solution for the detached mode 
appears as a blue solid curve in the top panel of Figure 3, and the corresponding light residuals are plotted in the middle 
panel of the figure. The semi-detached mode also led to the same results. In all procedures, we included the orbital 
eccentricity as an adjustable parameter, but its value remained indistinguishable from zero. This indicates that 
our program target is in a circular orbit.

\section{LIGHT-CURVE RESIDUALS AND PULSATIONAL CHARACTERISTICS}

From its temperature, the primary star of EPIC 245932119 would be a candidate for $\delta$ Sct and/or $\gamma$ Dor pulsators.
Pigulski \& Michalska (2007) reported that the EB system is pulsating at frequencies of $f_{\rm PM_1}$ = 26.16925 days$^{-1}$ 
and possibly $f_{\rm PM_2}$ = 23.0413 days$^{-1}$. So then, the pulsations can be attributed to $\delta$ Sct-type variability. 
In order to search for more reliable pulsation frequencies in the binary star, we followed an approach analogous to that of 
Lee  et al. (2017). First of all, we made a total of 46 light curves at intervals of one orbital period and 
separately analyzed them for the detached and semi-detached modes by adjusting only the reference epoch ($T_0$) among 
the photometric parameters in Table 1. Then, we performed a multiple frequency analysis for the corresponding light residuals 
in the outside-primary eclipsing phase. The PERIOD04 program (Lenz \& Breger 2005) was carried out in the frequency range from 
0 to the Nyquist limit of $f_{\rm Ny}$ = 24.47 days$^{-1}$. After the successive prewhitening of each frequency peak 
(see Lee et al. 2014), we detected the frequencies based on the empirical threshold of the signal to noise amplitude ratio 
(S/N) larger than 4.0 (Breger et al. 1993). Third, we removed the pulsation signatures from the observed {\it K{\rm 2}} data. 
The pulsation-subtracted data were modeled using the W-D code, and the new light-curve parameters were used to reanalyze 
the 46 light curves in the first step.

This procedure was iterated 7 and 5 times, respectively, for the detached and semi-detached modes until both 
detected frequencies and binary parameters were unchanged. The physical parameters of EPIC 245932119 are given in Table 2, 
and the pulsation-subtracted data and synthetic light curve for the detached mode are illustrated in the top panel of 
Figure 3. In Tables 1 and 2, the binary parameters from the observed and pulsation-subtracted data are in good agreement with 
each other. The light residuals after removing the binary effects from the observed data for the detached mode are plotted in 
Figure 4 as magnitude versus BJD, where the lower panel presents a short section of the residuals. As the result of 
our detailed analyses, we found 35 frequencies with the criterion of S/N $>$ 4.0. The periodogram from the PERIOD04 program 
is shown in Figure 5. The spectral window in the first panel shows strong side bands at integer multiples of 
the orbital frequency ($f_{\rm orb}$ = 0.62727 days$^{-1}$), which may be produced by excluding the data of the primary eclipses. 
The amplitude spectra for EPIC 245932119 before and after prewhitening the first 7 frequencies and then all 35 frequencies are 
shown in the second to bottom panels of Figure 5, respectively. The results are listed in Table 3, wherein the uncertainties 
were calculated according to Kallinger et al. (2008). The synthetic curve calculated from the 35-frequency fit is displayed in 
the lower panel of Figure 4. 

As shown in Figure 5 and Table 3, the main signals of EPIC 245932119 lie in two frequency regions of $<$ 7 days$^{-1}$ and 
$>$ 19 days$^{-1}$. Within the frequency resolution of 1.5/$\Delta T$ = 0.019 days$^{-1}$ ($\Delta T$ is the time base of 
observations; Loumos \& Deeming 1978), we searched for possible harmonic and combination frequencies. As a consequence, 
the six frequencies ($f_8$, $f_9$, $f_{14}$, $f_{18}$, $f_{24}$, $f_{32}$) in the low frequency region are 
the orbital frequency of $f_{\rm orb}$ and its multiples, which can be partially attributed to imperfect removal of 
the binary effects from the observed data. Further, $f_{19}$, $f_{20}$, and $f_{22}$ appear to be the sidelobes split 
from the $f_1$ frequency by $f_{\rm orb}$. On the other hand, some frequencies near the Nyquist limit $f_{\rm Ny}$ can be 
reflections of real pulsations (2$f_{\rm Ny}-f_i$) higher than $f_{\rm Ny}$ (Murphy et al. 2013; Lee et al. 2016b). It is 
possible that our frequency of $f_1$ = 22.77503 days$^{-1}$ may be a reflection (2$f_{\rm Ny}-f_{\rm PM_1}$ = 22.77075 days$^{-1}$) 
of a signal detected by Pigulski \& Michalska (2007). However, at present, we find it difficult to distinguish real peaks 
from the Nyquist aliases.

\section{DISCUSSION AND CONCLUSIONS}

In this paper, we have studied both binarity and pulsation of EPIC 245932119 from the {\it K{\rm 2}} data made during 
the Campaign 12. The binary light curve was satisfactorily analyzed for two cases: including and removing oscillation 
frequencies. The results indicate that the EB system is in a detached configuration with $q=0.283$ or 
a semi-detached configuration with $q=0.245$, and that its binary parameters are not affected by the pulsations. 
In the detached mode, both components fill $F_1=77$ \% and $F_2=93$ \% of their inner critical lobe, respectively, 
while $F_1=73$ \% in the semi-detached mode. Here, the fill-out factor $F_{1,2}=\Omega_{\rm in}/\Omega_{1,2}$, where 
$\Omega_{\rm in}$ is the potential of the inner Roche surface. There is currently no way to know which Roche configuration 
is more appropriate to describe our program target.

The effective temperature of the primary component corresponds to a normal main-sequence star with a spectral type of 
about A7V (Harmanec 1988; Pecaut \& Mamajek 2013). Based on the empirical relation between spectral type and stellar mass, 
we estimated the primary's mass to be $M_1$ = 1.75 $\pm$ 0.18 $M_\odot$ with an error of 10 \% assumed. Then, 
the absolute dimensions of EPIC 245932119 were computed from the light-curve parameters and the $M_1$ value, and they are 
listed in the bottom of Table 2. We calculated the luminosity ($L$) and bolometric magnitudes ($M_{\rm bol}$) by using 
$T_{\rm eff}$$_\odot$ = 5,780 K and $M_{\rm bol}$$_\odot$ = +4.73 for solar values. The bolometric corrections (BCs) 
were derived from the expression between $\log T_{\rm eff}$ and BC (Torres 2010). Using the estimated parameters, 
we examined the evolutionary state of EPIC 245932119 in mass-radius, mass-luminosity, and Hertzsprung-Russell (HR) diagrams 
(\. Ibano\v{g}lu et al. 2006; Lee \& Park 2018). In these diagrams, the pulsating primary star resides within 
the $\delta$ Sct instability strip on the main-sequence band, while the low-mass companion is remarkly larger and brighter 
than expected for main-sequence stars of the same mass. This indicates that the secondary component is highly evolved.

With an apparent visual magnitude of $V$ = +9.604 $\pm$ 0.035 (H\o g et al. 2000) and the interstellar reddening 
of $A_{\rm V}$ = 0.100 (Schlafly \& Finkbeiner 2011), we determined the distance of our program target to be 429 $\pm$ 23 pc 
and 412 $\pm$ 22 pc, respectively, for the detached and semi-detached modes. Considering that the physical properties of 
EPIC 245932119 were obtained from the mass-spectral type relation without the radial velocities of each component, 
these values are in satisfactory agreement with the Gaia distance of 463 $\pm$ 11 pc computed from 
Gaia DR2\footnote{https://gea.esac.esa.int/archive/} (2.16 $\pm$ 0.05 mas; Gaia Collaboration et al. 2018). On this account, 
it is not possible to distinguish between the two Roche modes from the distance determination. 

In order to investigate the pulsational characteristics of EPIC 245932119, multiple frequency analyses were applied to 
the light-curve residuals in the out-of-primary eclipsing phase after removing the binarity effects from the observed data. 
As a result, we found the 35 frequencies with S/N larger than 4.0 in two ranges of 0.62$-$6.28 days$^{-1}$ and 
19.36$-$24.07 days$^{-1}$. Among these, nine frequencies are orbital harmonics ($f_8$, $f_9$, $f_{14}$, $f_{18}$, 
$f_{24}$, $f_{32}$) and combination terms ($f_{19}$, $f_{20}$, $f_{22}$), most of which may arise from alias effects caused 
by the orbital frequency. Applying the absolute parameters in the detached mode of Table 2 to the well-known relation of 
$\log Q_i = -\log f_i + 0.5 \log g + 0.1M_{\rm bol} + \log T_{\rm eff} - 6.456$ (Petersen \& J\o rgensen 1972), we computed 
the pulsation constants for the remaining 26 frequencies, and they are given in the sixth column of Table 3. The $Q$ values 
of 0.013$-$0.016 days and the pulsation periods ($P_{\rm pul}$) of 0.041$-$0.052 days correspond to the low-order pressure 
modes of $\delta$ Sct pulsators with the typical ranges of $Q <$ 0.04 days and $P_{\rm pul}$ = 0.02$-$0.2 days (Breger 2000). 
Moreover, as mentioned above, the primary component lies inside the $\delta$ Sct region of the HR diagram. The results 
demonstrate that EPIC 245932119 is an EB system with a $\delta$ Sct-type pulsating component. The binary parameters in 
Table 2, the pulsation quantities in Table 3, and the gravitational force of $\log (F/M_1) =$ 2.76 
(that is applied to the pulsating primary by its companion) match well empirical relations of the $\delta$ Sct pulsators 
in binaries (Kahraman Ali\c cavu\c s et al. 2017; Liakos \& Niarchos 2017). 

Because EPIC 245932119 is bright enough for spectroscopic follow-up or other observations, future high-resolution 
spectra will assist in measuring the double-lined radial velocities and spectroscopic mass ratio and hence in determining 
its Roche configuration and absolute parameters. High-cadence multi-band photometry is needed to identify the detected 
frequencies and pulsation modes of the EB system. These offer us an important piece of observational evidence for studying 
the interesting objects.

\acknowledgments{ }
This paper includes data collected by the {\it K{\rm 2}} mission. Funding for the {\it K{\rm 2}} mission is provided by 
the NASA Science Mission directorate. Some of the data presented in this paper were obtained from the Mikulski Archive for 
Space Telescopes (MAST). We appreciate the careful reading and valuable comments of the anonymous referee. This research 
has made use of the Simbad database maintained at CDS, Strasbourg, France, and was supported by the KASI grant 2018-1-830-02. 
K.H. was supported by the grant numbers 2017R1A4A1015178 of the National Research Foundation (NRF) of Korea. 
M.H.K. acknowledges Allan R. Schmitt for making the LcTools software.

\clearpage
\begin{figure}
\includegraphics[scale=0.85]{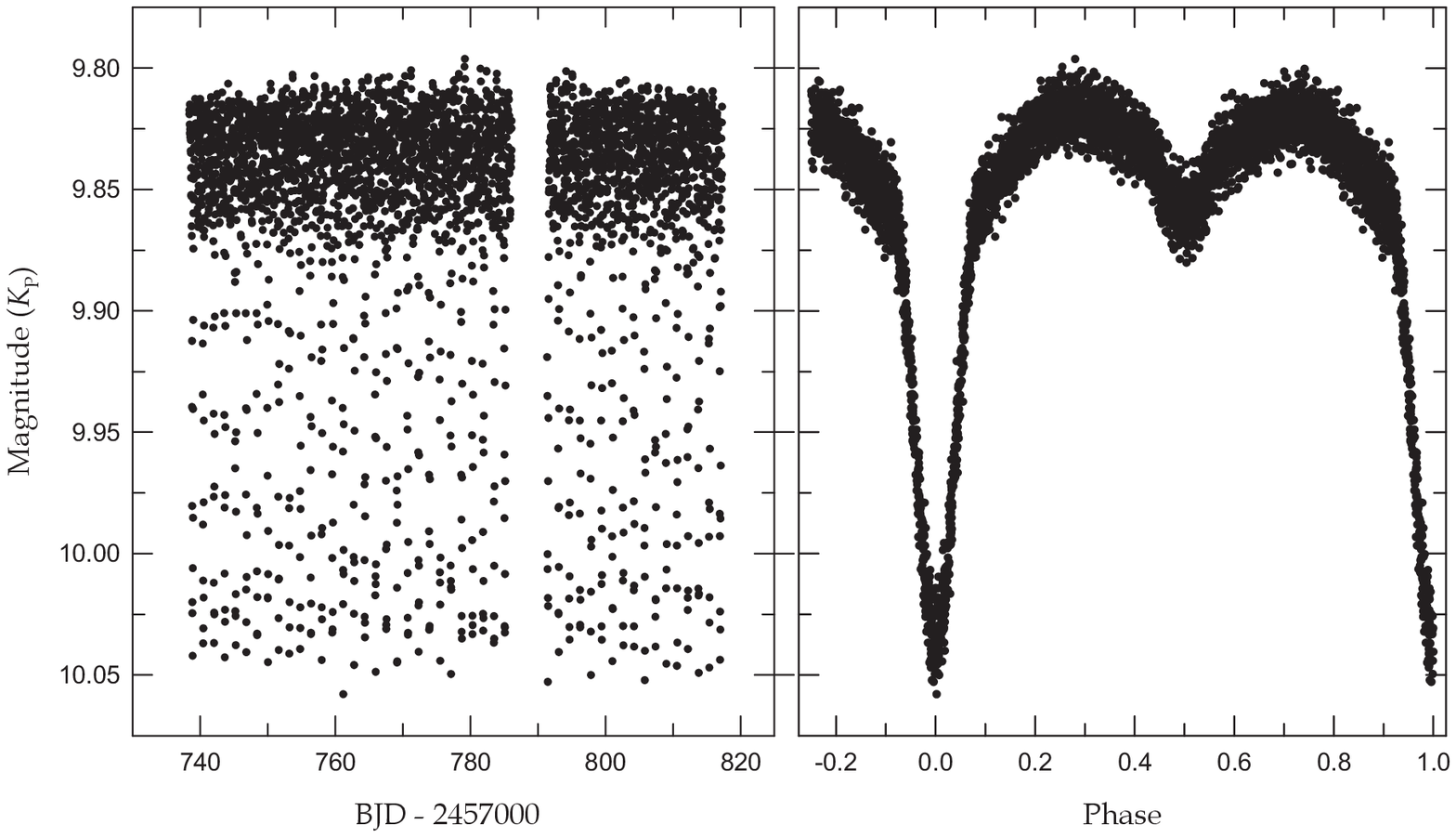}
\caption{{\it K{\rm 2}} observations of EPIC 245932119 distributed in BJD (left panel) and orbital phase (right panel). }
\label{Fig1}
\end{figure}

\begin{figure}
\includegraphics[]{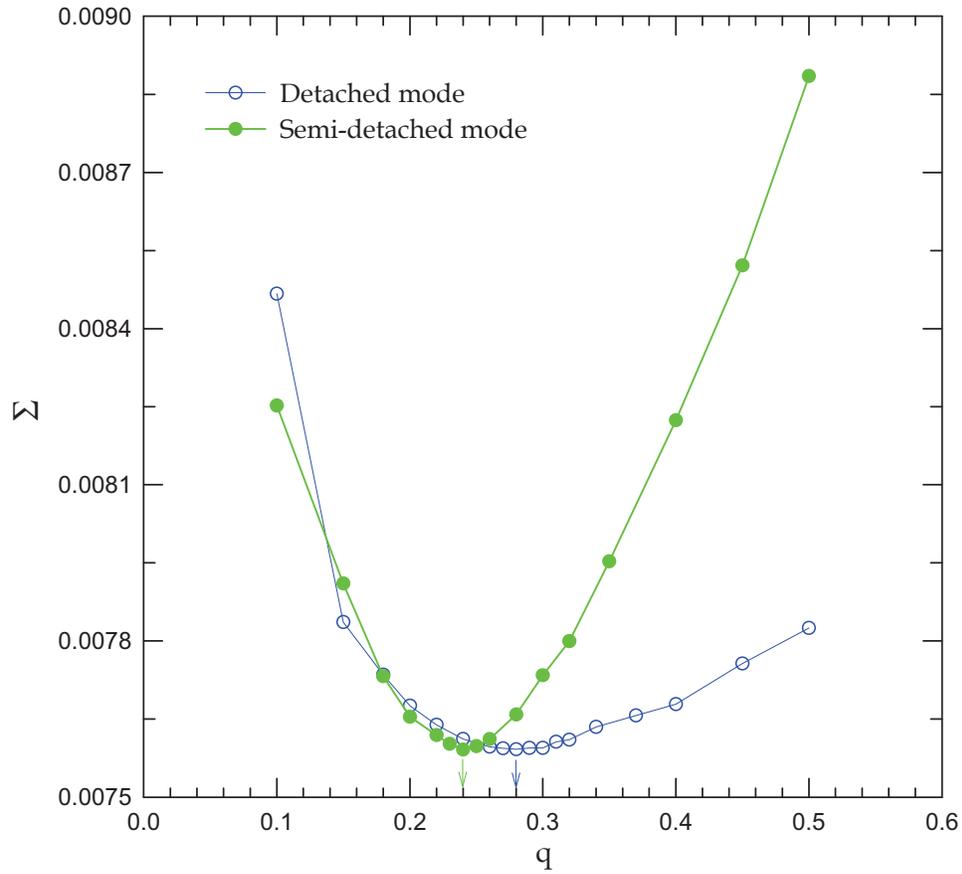}
\caption{Behavior of $\Sigma$ (the weighted sum of the residuals squared) as a function of the mass ratio $q$. 
The open and filled circles represent the $q$-search results for the detached and semi-detached configurations, respectively. }
\label{Fig2}
\end{figure}

\begin{figure}
\includegraphics[]{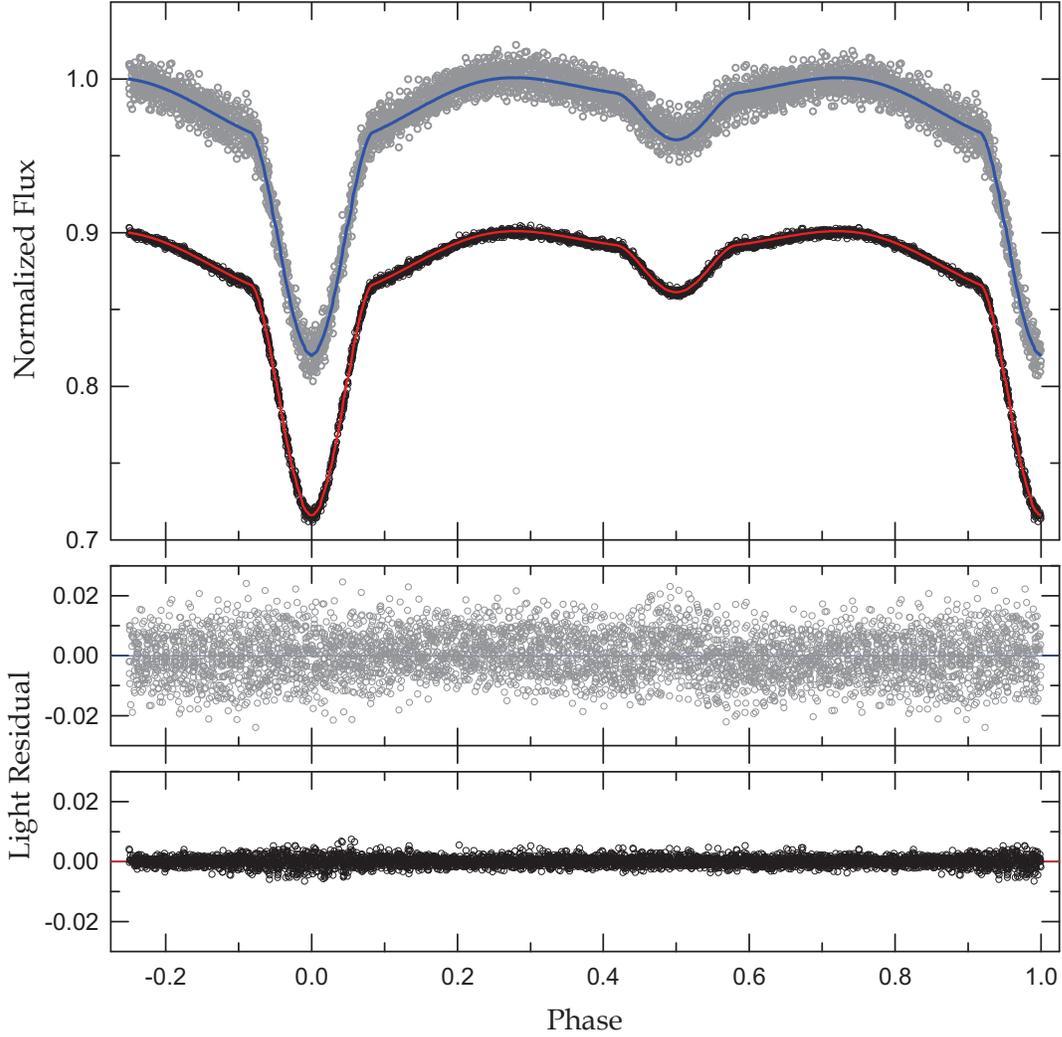}
\caption{Binary light curve before (gray circle) and after (black circle) subtracting the pulsation signatures from 
the observed {\it K{\rm 2}} data. The blue and red solid curves are computed with the photometric solutions for 
the detached modes in Tables 1 and 2, respectively. The corresponding residuals from the fits are plotted at the middle 
and bottom panels in the same order as the light curves. }
\label{Fig3}
\end{figure}

\begin{figure}
\includegraphics[]{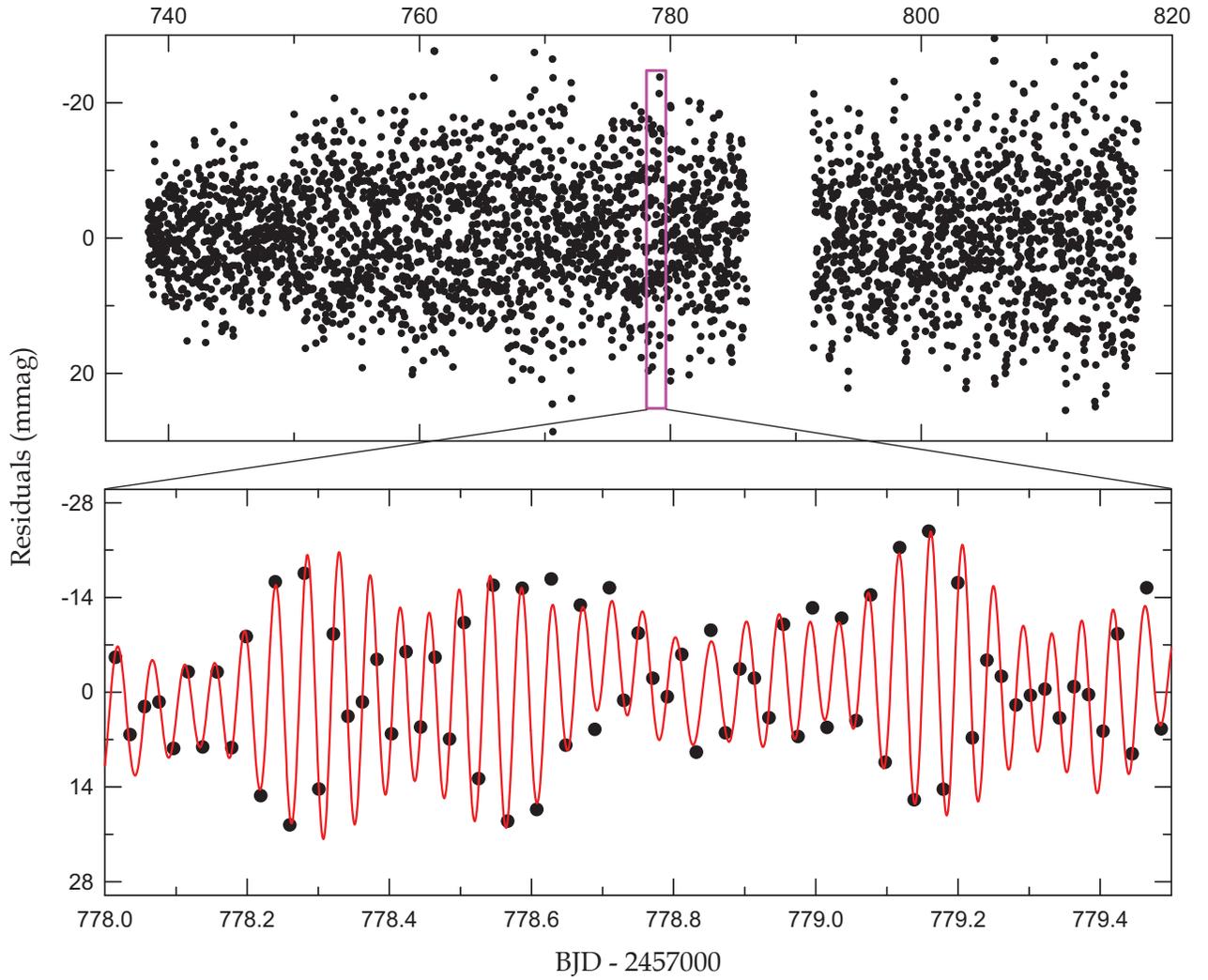}
\caption{Light-curve residuals after removing the binarity effects from the observed data. The lower panel presents 
a short section of the residuals marked by the inset box in the upper panel. The synthetic curve is computed from 
the 35-frequency fit to the data. }
\label{Fig4}
\end{figure}

\begin{figure}
\includegraphics[]{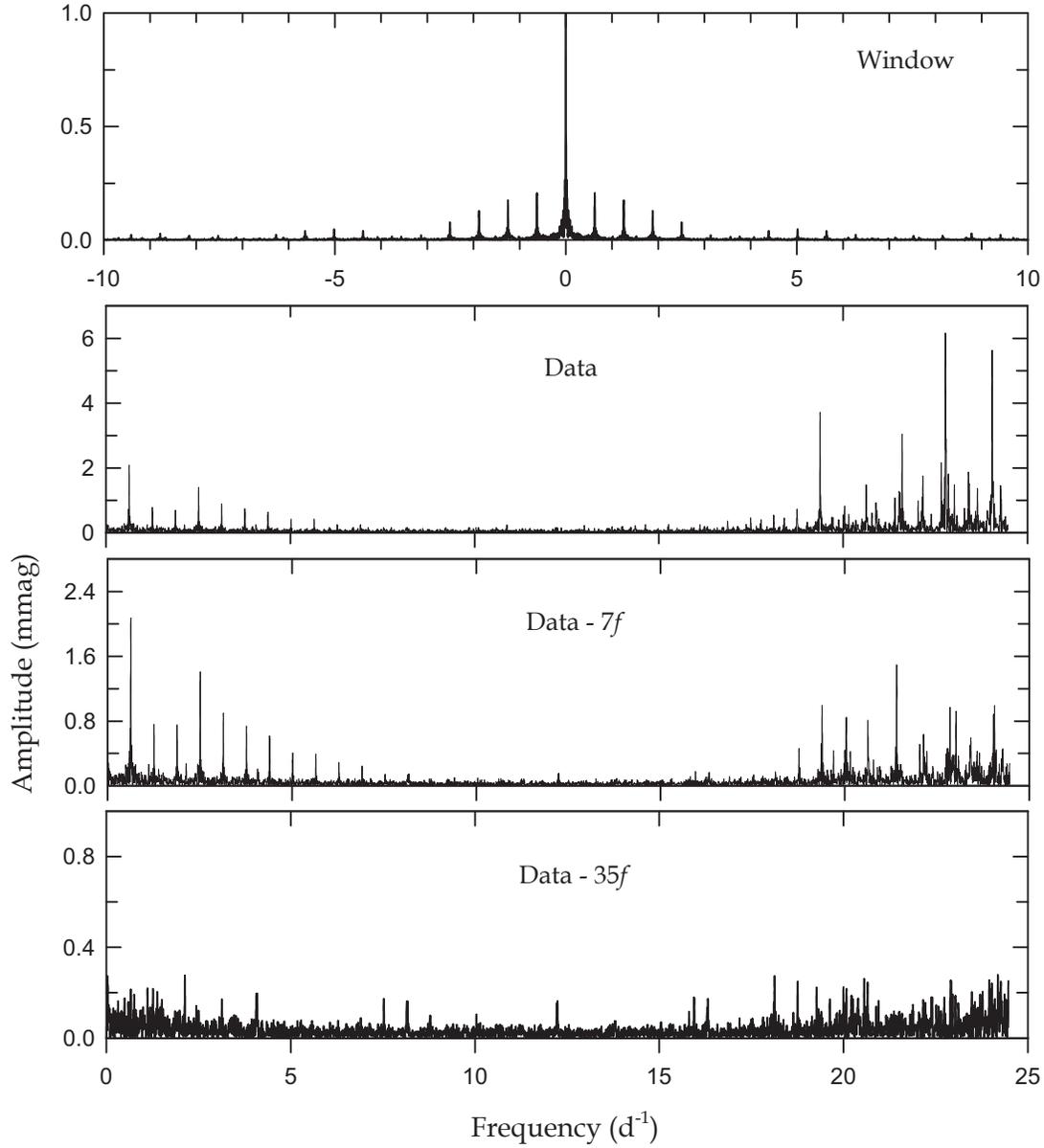}
\caption{Periodogram from the PERIOD04 program for the entire light residuals in the out-of-primary eclipsing phase 
(0.09$-$0.91$P$). The window spectrum is displayed in the top panel. The amplitude spectra before and after prewhitening 
the first 7 frequencies and all 35 frequencies are shown in the second to bottom panels. }
\label{Fig5}
\end{figure}

\clearpage
\begin{deluxetable}{lcccccccc}
\tablewidth{0pt} 
\tablecaption{Binary Parameters of EPIC 245932119 from the Observed Data}
\tablehead{
\colhead{Parameter}                      & \multicolumn{2}{c}{Detached Mode}           && \multicolumn{2}{c}{Semi-detached Mode}      \\ [1.0mm] \cline{2-3} \cline{5-6} \\[-2.0ex]
                                         & \colhead{Primary} & \colhead{Secondary}     && \colhead{Primary} & \colhead{Secondary}                                                  
}                                                                                                                                     
\startdata                                                                                                                            
$T_0$ (BJD)                              & \multicolumn{2}{c}{2,457,773.92915(35)}     && \multicolumn{2}{c}{2,457,773.92896(35)}     \\
$P$ (day)                                & \multicolumn{2}{c}{1.594195(23)}            && \multicolumn{2}{c}{1.594197(23)}            \\
$q$                                      & \multicolumn{2}{c}{0.2810(13)}              && \multicolumn{2}{c}{0.2411(15)}              \\
$i$ (deg)                                & \multicolumn{2}{c}{68.31(33)}               && \multicolumn{2}{c}{66.74(12)}              \\
$T$ (K)                                  & 7652(95)          & 3824(41)                && 7652(95)          & 4038(34)                \\
$\Omega$                                 & 3.119(13)         & 2.600(22)               && 3.180(26)         & 2.332                   \\
$\Omega_{\rm in}$                        & \multicolumn{2}{c}{2.424}                   && \multicolumn{2}{c}{2.332}                   \\
$A$                                      & 1.0               & 0.5                     && 1.0               & 0.5                     \\
$g$                                      & 1.0               & 0.32                    && 1.0               & 0.32                    \\
$X$, $Y$                                 & 0.672, 0.200      & 0.611, 0.156            && 0.672, 0.200      & 0.617, 0.151            \\
$x$, $y$                                 & 0.597, 0.239      & 0.745, 0.033            && 0.597, 0.239      & 0.754, 0.043            \\
$L$/($L_{1}$+$L_{2}$)                    & 0.9837(26)        & 0.0163                  && 0.9644(26)        & 0.0356                  \\
$r$ (pole)                               & 0.3505(17)        & 0.2223(37)              && 0.3388(45)        & 0.2456(5)               \\
$r$ (point)                              & 0.3722(23)        & 0.2521(66)              && 0.3554(56)        & 0.3585(6)               \\
$r$ (side)                               & 0.3609(20)        & 0.2281(42)              && 0.3475(50)        & 0.2556(5)               \\
$r$ (back)                               & 0.3672(21)        & 0.2434(54)              && 0.3521(53)        & 0.2882(5)               \\
$r$ (volume)$\rm ^a$                     & 0.3597(20)        & 0.2315(49)              && 0.3462(51)        & 0.2640(5)               \\ 
$\sum W(O-C)^2$                          & \multicolumn{2}{c}{0.00756}                 && \multicolumn{2}{c}{0.00756}                 \\ 
\enddata
\tablenotetext{a}{Mean volume radius.}
\end{deluxetable}

\clearpage
\begin{deluxetable}{lcccccccc}
\tablewidth{0pt} 
\tablecaption{Binary Parameters of EPIC 245932119 from the Pulsation-Subtracted Data}
\tablehead{
\colhead{Parameter}                      & \multicolumn{2}{c}{Detached Mode}           && \multicolumn{2}{c}{Semi-detached Mode}      \\ [1.0mm] \cline{2-3} \cline{5-6} \\[-2.0ex]
                                         & \colhead{Primary} & \colhead{Secondary}     && \colhead{Primary} & \colhead{Secondary}                                                  
}                                                                                                                                     
\startdata                                                                                                                            
$T_0$ (BJD)                              & \multicolumn{2}{c}{2,457,773.929462(70)}    && \multicolumn{2}{c}{2,457,773.929566(73)}    \\
$P$ (day)                                & \multicolumn{2}{c}{1.5941988(47)}           && \multicolumn{2}{c}{1.5942005(49)}           \\
$q$                                      & \multicolumn{2}{c}{0.2825(13)}              && \multicolumn{2}{c}{0.2450(16)}              \\
$i$ (deg)                                & \multicolumn{2}{c}{68.69(13)}               && \multicolumn{2}{c}{66.75(7)}                \\
$T$ (K)                                  & 7652(95)          & 3834(30)                && 7652(95)          & 4039(12)                \\
$\Omega$                                 & 3.155(16)         & 2.609(8)                && 3.212(19)         & 2.341                   \\
$\Omega_{\rm in}$                        & \multicolumn{2}{c}{2.427}                   && \multicolumn{2}{c}{2.341}                   \\
$X$, $Y$                                 & 0.672, 0.200      & 0.611, 0.156            && 0.672, 0.200      & 0.617, 0.151            \\
$x$, $y$                                 & 0.597, 0.239      & 0.746, 0.035            && 0.597, 0.239      & 0.754, 0.043            \\
$L$/($L_{1}$+$L_{2}$)                    & 0.9829(10)        & 0.0171                  && 0.9633(10)        & 0.0367                  \\
$r$ (pole)                               & 0.3462(19)        & 0.2218(22)              && 0.3356(23)        & 0.2468(5)               \\
$r$ (point)                              & 0.3667(25)        & 0.2510(37)              && 0.3517(29)        & 0.3601(6)               \\
$r$ (side)                               & 0.3561(22)        & 0.2275(24)              && 0.3440(26)        & 0.2568(5)               \\
$r$ (back)                               & 0.3621(24)        & 0.2425(30)              && 0.3485(27)        & 0.2894(5)               \\
$r$ (volume)                             & 0.3550(22)        & 0.2309(28)              && 0.3428(26)        & 0.2652(5)               \\ 
$\sum W(O-C)^2$                          & \multicolumn{2}{c}{0.00153}                 && \multicolumn{2}{c}{0.00155}                 \\ [1.0mm]
\multicolumn{6}{l}{Absolute parameters:}                                                                                              \\            
$M$ ($M_\odot$)                          & 1.75(18)          & 0.49(5)                 && 1.75(18)          & 0.43(4)                 \\
$R$ ($R_\odot$)                          & 2.67(12)          & 1.73(8)                 && 2.55(11)          & 1.97(9)                 \\
$\log$ $g$ (cgs)                         & 3.83(6)           & 3.65(6)                 && 3.87(6)           & 3.48(6)                 \\
$L$ ($L_\odot$)                          & 22(2)             & 0.58(6)                 && 20(2)             & 0.93(8)                 \\
$M_{\rm bol}$ (mag)                      & 1.38(11)          & 5.32(11)                && 1.48(11)          & 4.81(10)                \\
BC (mag)                                 & 0.03(1)           & $-$1.41(6)              && 0.03(1)           & $-$1.07(2)              \\
$M_{\rm V}$ (mag)                        & 1.35(11)          & 6.73(12)                && 1.45(11)          & 5.88(10)                \\          
Distance (pc)                            & \multicolumn{2}{c}{429$\pm$23}              && \multicolumn{2}{c}{412$\pm$22}              \\
\enddata
\end{deluxetable}

\begin{deluxetable}{lrcccccc}
\tabletypesize{\small}  
\tablewidth{0pt}
\tablecaption{Multiple Frequency Analysis of EPIC 245932119$\rm ^a$ }
\tablehead{
             & \colhead{Frequency}    & \colhead{Amplitude} & \colhead{Phase} & \colhead{S/N$\rm ^b$} & \colhead{$Q$}    & \colhead{Remark}      \\
             & \colhead{(day$^{-1}$)} & \colhead{(mmag)}    & \colhead{(rad)} &                       & \colhead{(days)} &
}                                                                                                                        
\startdata                                                                                                               
$f_{1}$      & 22.77503$\pm$0.00005   & 6.42$\pm$0.12       & 1.94$\pm$0.06   & 88.62                 & 0.013            &                       \\
$f_{2}$      & 24.04266$\pm$0.00005   & 5.86$\pm$0.13       & 0.42$\pm$0.06   & 78.88                 & 0.013            &                       \\
$f_{3}$      & 19.37332$\pm$0.00006   & 3.43$\pm$0.08       & 2.79$\pm$0.07   & 70.71                 & 0.016            &                       \\
$f_{4}$      & 22.79150$\pm$0.00009   & 3.62$\pm$0.12       & 0.30$\pm$0.10   & 49.92                 & 0.013            &                       \\
$f_{5}$      & 21.60056$\pm$0.00010   & 2.71$\pm$0.11       & 3.93$\pm$0.12   & 42.91                 & 0.014            &                       \\
$f_{6}$      & 24.05154$\pm$0.00010   & 3.24$\pm$0.13       & 5.07$\pm$0.11   & 43.64                 & 0.013            &                       \\
$f_{7}$      & 22.66157$\pm$0.00013   & 2.40$\pm$0.12       & 1.24$\pm$0.15   & 33.22                 & 0.013            &                       \\
$f_{8}$      &  0.62685$\pm$0.00013   & 2.09$\pm$0.11       & 5.88$\pm$0.16   & 32.22                 &                  & $f_{\rm orb}$         \\
$f_{9}$      &  2.50992$\pm$0.00013   & 1.78$\pm$0.09       & 5.84$\pm$0.15   & 33.81                 &                  & 4$f_{\rm orb}$        \\
$f_{10}$     & 21.40661$\pm$0.00017   & 1.55$\pm$0.10       & 0.74$\pm$0.20   & 25.67                 & 0.014            &                       \\
$f_{11}$     & 19.38409$\pm$0.00022   & 0.96$\pm$0.08       & 1.52$\pm$0.25   & 19.66                 & 0.016            &                       \\
$f_{12}$     & 24.06168$\pm$0.00032   & 1.00$\pm$0.13       & 4.43$\pm$0.38   & 13.35                 & 0.013            &                       \\
$f_{13}$     & 22.85552$\pm$0.00031   & 1.00$\pm$0.12       & 1.24$\pm$0.36   & 13.76                 & 0.013            &                       \\
$f_{14}$     &  3.76425$\pm$0.00015   & 1.18$\pm$0.07       & 5.06$\pm$0.17   & 28.75                 &                  & 6$f_{\rm orb}$        \\
$f_{15}$     & 23.01905$\pm$0.00035   & 0.90$\pm$0.13       & 0.96$\pm$0.41   & 12.20                 & 0.013            &                       \\
$f_{16}$     & 20.04010$\pm$0.00028   & 0.82$\pm$0.09       & 5.20$\pm$0.32   & 15.49                 & 0.015            &                       \\
$f_{17}$     & 20.62638$\pm$0.00029   & 0.84$\pm$0.10       & 2.43$\pm$0.34   & 14.87                 & 0.015            &                       \\
$f_{18}$     &  5.01858$\pm$0.00018   & 0.79$\pm$0.06       & 4.20$\pm$0.21   & 23.92                 &                  & 8$f_{\rm orb}$        \\
$f_{19}$     & 22.13804$\pm$0.00049   & 0.61$\pm$0.12       & 2.78$\pm$0.56   &  8.90                 &                  & $f_1-f_{\rm orb}$     \\
$f_{20}$     & 23.41835$\pm$0.00041   & 0.77$\pm$0.12       & 2.49$\pm$0.47   & 10.60                 &                  & $f_1+f_{\rm orb}$     \\
$f_{21}$     & 24.03189$\pm$0.00043   & 0.75$\pm$0.13       & 3.79$\pm$0.50   & 10.11                 & 0.013            &                       \\
$f_{22}$     & 23.40060$\pm$0.00046   & 0.68$\pm$0.12       & 3.95$\pm$0.53   &  9.37                 &                  & $f_1+f_{\rm orb}$     \\
$f_{23}$     & 22.78326$\pm$0.00040   & 0.79$\pm$0.12       & 3.93$\pm$0.46   & 10.89                 & 0.013            &                       \\
$f_{24}$     &  6.27291$\pm$0.00029   & 0.46$\pm$0.05       & 3.41$\pm$0.33   & 15.09                 &                  & 10$f_{\rm orb}$       \\
$f_{25}$     & 19.69403$\pm$0.00051   & 0.43$\pm$0.09       & 1.87$\pm$0.59   &  8.43                 & 0.015            &                       \\
$f_{26}$     & 23.58885$\pm$0.00073   & 0.44$\pm$0.13       & 6.14$\pm$0.85   &  5.91                 & 0.013            &                       \\
$f_{27}$     & 23.08496$\pm$0.00076   & 0.41$\pm$0.12       & 3.54$\pm$0.88   &  5.69                 & 0.013            &                       \\
$f_{28}$     & 22.93602$\pm$0.00082   & 0.38$\pm$0.12       & 3.98$\pm$0.95   &  5.27                 & 0.013            &                       \\
$f_{29}$     & 23.50772$\pm$0.00080   & 0.39$\pm$0.12       & 1.51$\pm$0.93   &  5.38                 & 0.013            &                       \\
$f_{30}$     & 23.44814$\pm$0.00068   & 0.46$\pm$0.12       & 5.59$\pm$0.79   &  6.38                 & 0.013            &                       \\
$f_{31}$     & 22.81622$\pm$0.00072   & 0.43$\pm$0.12       & 5.59$\pm$0.84   &  5.96                 & 0.013            &                       \\
$f_{32}$     &  0.63509$\pm$0.00076   & 0.37$\pm$0.11       & 2.40$\pm$0.88   &  5.70                 &                  & $f_{\rm orb}$         \\
$f_{33}$     & 23.91844$\pm$0.00094   & 0.34$\pm$0.13       & 0.23$\pm$1.09   &  4.61                 & 0.013            &                       \\
$f_{34}$     & 19.36318$\pm$0.00058   & 0.36$\pm$0.08       & 5.81$\pm$0.67   &  7.47                 & 0.016            &                       \\
$f_{35}$     & 22.74840$\pm$0.00097   & 0.32$\pm$0.12       & 4.56$\pm$1.13   &  4.43                 & 0.013            &                       \\
\enddata                                                                                                                           
\tablenotetext{a}{Frequencies are listed in order of detection. }
\tablenotetext{b}{Calculated in a range of 5 d$^{-1}$ around each frequency. }
\end{deluxetable}

\end{document}